\documentclass[aps,prd,twocolumn,superscriptaddress,preprintnumbers,showpacs]{revtex4-1}
\usepackage{amsmath,amsfonts,amssymb,graphicx,color,times,bbm,psfrag,dsfont,
slashed,bigints}
\usepackage[colorlinks=true,citecolor=blue]{hyperref}

\newcommand{\be}{\begin{equation}}
\newcommand{\ee}{\end{equation}}
\newcommand{\bal}{\begin{align}}
\newcommand{\eal}{\end{align}}

\newcommand{\red}{\textcolor{black}}

\usepackage{bm}

\usepackage[T1]{fontenc}
\usepackage[latin9]{inputenc}

\newcommand{\beq}{\begin{equation}}
\newcommand{\eeq}{\end{equation}}
\newcommand{\bea}{\begin{eqnarray}}
\newcommand{\eea}{\end{eqnarray}}

\begin{document}

\title{Gravitational wave echoes from strange stars}
\author{M. Mannarelli}
\email{massimo@lngs.infn.it}
\affiliation{INFN, Laboratori Nazionali del Gran Sasso, Via G. Acitelli,
22, I-67100 Assergi (AQ), Italy.}
%\author{G. Pagliaroli}
%\affiliation{Gran Sasso Science Institute, L'Aquila (AQ), Italy}
\author{F. Tonelli}
\email{francesco.tonelli@lngs.infn.it}
\affiliation{INFN, Laboratori Nazionali del Gran Sasso, Via G. Acitelli,
22, I-67100 Assergi (AQ), Italy.}
\affiliation{University of L'Aquila, Physics and Chemistry Department, L'Aquila, Italy}

\begin{abstract}
It has recently been claimed, with a $4.2 \sigma$ significance level, that gravitational wave echoes at a frequency of about $72$ Hz have been produced in the GW170817 event. The merging of compact stars can lead to the emission of gravitational waves echoes if the post-merger object features a photon-sphere capable of partially trapping the gravitational waves. If the post-merger source is a black hole,  a second internal reflection surface,  associated to quantum effects near the black hole horizon, must be present to avoid the gravitational wave capture. Alternatively, gravitational wave echoes  can be produced by  ultracompact stars crossing the  photon-sphere line in the mass-radius diagram during the neutron star merging. In this case, the second reflection surface is not needed. A recently proposed preliminary analysis using an incompressible \red{(and so unphysical)} equation of state suggests that gravitational wave echoes at a frequency of tens of Hz can be produced by an ultracompact star.  Since strange stars are extremely compact, we examine the possibility that strange stars emit gravitational wave echoes at such a frequency. Using \red{parameterized models of the equation of state of} ultra-stiff quark matter  we find that a strange star can emit gravitational wave echoes, but the corresponding frequencies are of the order of tens of  kHz, thus not compatible with the $72$ Hz signal.  
\end{abstract}


\maketitle
\section{Introduction} The intriguing possibility that the merging of  compact massive objects can lead to the emission of gravitational wave (GW) echoes, eventually detectable by the LIGO-VIRGO interferometers, has been investigated by various authors~\cite{Abedi:2016hgu, Abedi:2017isz, Abedi:2018npz}, but remains a controversial topic, see for example~\cite{Ashton:2016xff, Westerweck:2017hus}. The emission mechanism of GW echoes relies on the existence of a very massive post-merger object of mass $M$, featuring  a photon-sphere, see~\cite{Weinberg:1972kfs, Misner:1974qy, Claudel:2000yi, Virbhadra:1999nm}, leading to partial  GWs  trapping. The photon-sphere is a surface located at $R= 3M$ where circular photon orbits are possible thanks to an  angular potential barrier.  It is featured by both black holes, see for example the discussion in~\cite{Shapiro:1983du}, and  by ultracompact stars~\cite{1985CQGra...2..219I, Nemiroff:1993zz}.

For black holes  (BHs), GW echoes require a second reflection surface  to avoid the GWs absorption, related to quantum effects close to the BH horizon, see for example~\cite{Barcelo:2017lnx}.   As discussed in~\cite{Ferrari:2000sr},  GW echoes can also be produced by   ultracompact stars  featuring a photon-sphere. In this case, there is no need of an internal reflection surface  because, unlike BHs, the ultracompact star is not capable of absorbing a sizable fraction of GWs.

The GW170817  event~\cite{TheLIGOScientific:2017qsa} has been interpreted as the merging of two neutron stars (NSs)   with an estimated total mass  $M \approx 2.7 M_\odot$. The final stellar object has not been firmly established: it  can be  a  massive compact star or a BH. The possible presence of GW echoes in the GW170817 event has been analyzed in \cite{Abedi:2018npz}, where it is claimed that a signal at a frequency $\approx 72 $ Hz with a  4.2$\sigma$ significance level is present. The authors interpret this signal as originating from  quantum effects close to the BH horizon. 
An  interpretation of this echo signal as originating from an ultracompact star  has been first proposed in~\cite{Pani:2018flj}. This preliminary analysis, conducted by a simplified incompressible EoS, has  shown that  to produce  a signal at such a low frequency the stellar object formed in the coalescence of the NSs should be very compact,  close to the Buchdahl's limit radius~\cite{Buchdahl:1959zz}   $R_B = 9/4 M$. Thus, the compact stellar object  produced in the NS merging should have a compactness $M/R$ larger than $1/3$ to have a photon-sphere,  and smaller (but very close) to $4/9$ to emit GW echoes at a frequency of tens of Hz.

Since strange stars are known to be very compact~\cite{Alcock:1986hz,Haensel:1986qb},  we examine the possibility that the ultracompact object produced in the GW170817 event is a strange star and evaluate the frequency of the corresponding GW echoes. In particular, we  study  whether strange stars may have a photon-sphere and approach the  Buchdahl's limit. In our approach we assume that  the  conversion of nuclear matter to deconfined quark matter happens   by means of the  extremely high densities produced in the NS merging. An important aspect is, indeed,  that the analysis of the GW170817 tidal deformability   suggests that the   EoS of the merging NSs cannot be too stiff~\cite{TheLIGOScientific:2017qsa, Annala:2017llu, Most:2018hfd,Lim:2018bkq}, see also~\cite{Radice:2017lry} for an analysis based on multimessanger observations. Thus, the merging stellar objects can well be two standard NSs, or a   NS and a hybrid star~\cite{Nandi:2017rhy, Burgio:2018yix}, characterized by a not-too-stiff EoS. However, if the final stellar object emits GW echoes it has to be very compact and therefore with a different, very stiff EoS. For this reason we  assume that the source of the GW echoes  is a strange star \red{produced by the merging of the two NSs.} To have the most compact configuration we assume a simple  MIT bag model~\cite{Farhi:1984qu} EoS with the largest possible stiffness, corresponding to a speed of sound equal to the speed of light.

  \red{The formation of a strange star would certainly be accompanied by a release of energy, as discussed in framework of supernova explosions, see for example~\cite{Benvenuto:1989qr, Pagliara:2013tza, Ouyed:2017nuy}, possibly affecting the gamma and neutrino emissions associated to the merging of NSs.  The GW post merger emission could also be different, but we are not aware of any simulation of merging of NSs leading to the formation of a strange star. In the present paper we limit our  analysis to the post merger GW echo signal.}

 Although the strange star is initially hot and presumably in an highly excited state, possibly rotating at high frequency, we neglect both the temperature and the spinning effects, considering a static configuration of cold quark matter. We will then argue that both effects should be negligible in the present context. However, it is maybe of interest the fact that the excited strange star could relax also  emitting radio waves at kHz frequencies (or smaller)~\cite{Mannarelli:2014ija, Mannarelli:2015jia, Flores:2017kte}.

The present  study could, in principle, lead to interesting information on the quark matter EoS and on the possible realization of the  Bodmer and Witten   hypothesis~\cite{Bodmer:1971we,Witten:1984rs} that standard nuclei are not the ground state of matter. We remark that  although the  current astrophysical observations of masses and radii of NSs can in principle constrain the EoS of matter at supra-saturation densities, simultaneous mass and radius observations are difficult, meaning that several model EoSs, obtained considering rather different matter composition and  interactions, are capable of describing a wealth of astrophysical data.  The observation of NSs with  a gravitational mass $M\simeq 2 M_\odot$~\cite{Demorest:2010bx,Antoniadis:2013pzd} has challenged nuclear  EoSs, excluding the too soft ones. If a  compact star with an even larger mass, say of about $2.5 M_\odot$, is the final stellar object resulting in the NSs merging associated to the GW170817 event, although still compatible with extreme nuclear matter EoSs, it would certainly exclude a larger number of models, possibly challenging the present understanding of core-collapse neutron star formation~\cite{Lattimer:2012nd}.  As we will see, requiring that this compact object  emits  GW echoes   further constrains  the model EoSs, excluding the known nuclear EoSs, as already shown in~\cite{1985CQGra...2..219I,Pani:2018flj}, and constraining the quark matter EoS to be very stiff. Actually, even considering  extreme strange star models with a very stiff quark matter EoS we  can only marginally cross the photon-sphere radius line, obtaining  GW echoes frequencies of the order of tens of kHz. 

The present paper is organized as follows. In Sec.~\ref{sec:Model} we discuss the strange star model and obtain the corresponding mass-radius diagram, comparing strange stars with nuclear EoSs. In Sec.~\ref{sec:Frequency} we evaluate the typical GW echo frequency emitted by  the last stable strange star configuration. We draw our conclusions in Sec.~\ref{sec:Conclusions}. We use geometrized units, with $G=c=1$.

%  On the other hand, also NSs with estimated masses $\lesssim 1 M_\odot$ have been observed, see for example~\cite{Ozel:2012ax}, challenging the present understanding of core-collapse neutron star formation~\cite{Lattimer:2012nd}. 

\section{ The model}\label{sec:Model}
\begin{figure}
  \includegraphics[width=0.48\textwidth]{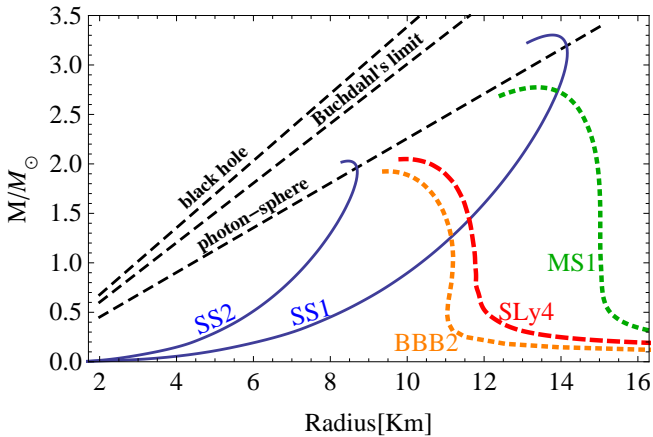}
\caption{Mass-radius diagram for various compact star models. The emission of GW echoes can only happen for those stellar models that cross the photon-sphere line. Standard NSs do not seem to be possible candidates. Strange stars with a maximally stiff EoS are  marginally compatible with this requirement.    }
\label{fig:MvsR}       % Give a unique label
\end{figure}

We  consider a simple bag model  EoS with energy density
\be\label{eq:simpleEoS}
\rho= p + 4 B\,,
\ee
where $p$ is the pressure, $B$ is the bag constant and the speed of sound has been set equal to $1$.  For simplicity we neglect the stellar rotation, thus the structure can be obtained solving the equations  of  Tolman-Oppenheimer-Volkov (TOV)  
\begin{align}\label{eq:TOV0}
\frac{ d \Phi}{dr} & =-\frac{1}{\rho+p} \frac{d p}{d r}\,, \\
\frac{d m}{d r}& = 4 \pi  \rho r^2\,,\label{eq:dm}\\
\frac{d p}{d r} &= (\rho+p)  \frac{m + 4 \pi p r^3}{2 m r -r^2} \label{eq:dp}\,,
\end{align}
where  $m(r)$ is the gravitational mass within the radius $r$ and $\Phi(r)$ is the gravitational potential.
The first equation follows from hydrostatic equilibrium and can be used to determine  the gravitational field inside the star once the pressure, and hence  the energy density, has been determined by solving Eqs.~\eqref{eq:dm} and~\eqref{eq:dp} iteratively. In Fig.~\ref{fig:MvsR} we report the obtained masses and radii  for two different values of the bag constant: $B_1=(145 \text{ MeV})^4$ (a typical bag model value) and  $B_2=(185 \text{  MeV})^4$, corresponding to the curves SS1 and SS2, respectively. With this extreme EoS, the $M(R)$ curves  cross the  photon-sphere line $M= R/3$, but do not approach the  Buchdahl's limit line.   The reason is that for small masses and radii, the stellar mass  is expected to grow as $R^3$, 
%\be
%M\propto R^3
%\ee
because strange quark matter is self-bound. Therefore, for small radii the $M(R) $ curve of strange stars stands below the photon-sphere radius. It can only approach it when the $M(R) $ curve bends, which happens for sufficiently large masses. For large masses the gravitational pull helps to  compress the structure,  however it eventually leads to an unstable branch, when a central density increase leads to a  gravitational mass reduction~\cite{Shapiro:1983du}. The last stable configurations, with the largest masses,  correspond to the tips of the $M(R)$ curves in the mass-radius diagram of Fig.~\ref{fig:MvsR}. These are as well the stable most compact configurations. Thus, it seems that strange stars cannot reach the  Buchdahl's limit line. The considered values of the bag constant lead to maximum masses  $M_\text{max}\approx 2  M_\odot$, for SS2, and of   $M_\text{max}\approx 3.3  M_\odot$ for SS1. Intermediate maximum masses can be obtained for values of the bag constant  in the range $B_1<B<B_2$, which can be easily inferred  considering that the maximum mass scales as~\cite{Witten:1984rs}
\be
M_\text{max} \propto B^{-1/2} \,.\ee Thus,  for values of the  bag constant in the above range, one spans  maximum masses   compatible with  the $2  M_\odot$ observations~\cite{Demorest:2010bx,Antoniadis:2013pzd}  and the GW170817 estimated total mass of $2.7 M_\odot$ \cite{TheLIGOScientific:2017qsa}. To make clear how extreme are these cases, consider that the central baryonic densities of these strange stars are about $25$ times the nuclear saturation density. Actually, such extreme  values of the baryonic densities are in agreement with the results obtained by simple models of NS collapse~\cite{1991A&A...252..651G} and by numerical simulations including rotation, see for example~\cite{Shibata:2003iw, Baiotti:2004wn}. In these works,  polytropic  EoSs are used to mimic nuclear matter.  Instead, in our approach we assume, maybe more reasonably, that at such large densities quark matter is liberated~\cite{Cabibbo:1975ig} and thus  the collapse of two NSs leads to the formation of a strange star. Whether the strange star is the final stellar object or it collapses to a black hole depends, in our very simple model,  on the value of the bag constant. Small values of the bag constant do indeed allow to have strange stars with a large mass. Hereafter we assume that the final stellar object is a strange star, but we will comment on the possible collapse of a strange star to a black hole.

One may expect that a different quark matter EoS   could provide a structure approaching the  Buchdahl's limit line in Fig.~\ref{fig:MvsR}.  A very general parameterization of the quark matter EoS is~\cite{Alford:2004pf}
\be
P =  \frac{3}{4 \pi^2} a_4 \mu^4 - \frac{3}{4 \pi^2}  a_2 \mu^2 - B\,,
\label{eq:EoS_quark_matter}
\ee
where $a_4$, $a_2$ are parameters  independent of the average quark chemical potential $\mu$. Varying these parameters  we obtain  last stable strange stars that are less compact than those reported in Fig.~\ref{fig:MvsR},  basically because   the  EoS in Eq.~\eqref{eq:EoS_quark_matter} is less stiff than the simple parameterization in Eq~\eqref{eq:simpleEoS}. See for example the  mass-radius diagram reported in~\cite{Mannarelli:2014ija} for some $M(R)$ results obtained with the parameterization in Eq~\eqref{eq:EoS_quark_matter}.

Regarding standard nuclear matter, as already noted in~\cite{1985CQGra...2..219I,Pani:2018flj},  the $M(R)$ curves obtained by the nuclear EoSs  approach the photon-sphere line from below, but do not cross it. As representative examples we consider in Fig~\ref{fig:MvsR}  the  BBB2 \cite{Baldo:1997ag}, the SLy4 \cite{Douchin:2001sv} and the MS1~\cite{Mueller:1996pm} EoSs, which at the largest possible mass values  have a speed of sound in the central region  close to $1$, but nonetheless are not sufficiently compact to cross the photon-sphere line. 

\section{Frequency of the gravitational wave echoes}\label{sec:Frequency}
In the proposed model the GWs emitted by the stellar object are partially reflected back by the angular potential barrier at the photon-sphere. \red{One may  indeed conceive  the photon-sphere as a trap for GWs, with characteristic  frequencies of the order of the inverse of the  length scale of the trap}. Thus, the smaller is the trap, {\it i.e} the closer is the stellar solution to the photon-sphere line in Fig.~\ref{fig:MvsR}, the larger is the GW echo frequency. Even considering  the last stable strange stars, corresponding to the most compact configuration, we obtain solutions that do not approach the Buchdahl's limit. For this reason  we expect that only GW echoes at large frequencies are produced. 

The   typical echo time can be evaluated as the light time from the center of the star to the photon-sphere, see~\cite{Pani:2018flj},  corresponding to 
\be\label{eq:tau_echo}
\tau_\text{echo} = \bigintss_0^{3M} \hspace{-.5cm}\frac{dr}{\sqrt{e^{2\Phi (r)}\left(1- \frac{2 m(r)}{r}\right)}}\,,
\ee
where the $m(r)$ and $\Phi(r)$ are  determined by solving the TOV's equations in Eqs.~(\ref{eq:TOV0}-\ref{eq:dp}).  We are assuming, quite reasonably, that GWs are not absorbed by the strange star. The GW echo frequency can be approximated by $\omega_\text{echo}=\pi/\tau_\text{echo}$~\cite{Cardoso:2016rao,Cardoso:2016oxy,Cardoso:2017cqb,Cardoso:2017njb,Mark:2017dnq}. \red{In~\cite{Abedi:2018npz} the estimated  frequency is given by $1/(2 \tau_\text{echo})$, which should actually  correspond to the repetition frequency of the echo signal. The  argument underlying our approximation is that the echo frequency corresponds to that of standing waves inside the photon-sphere,  see for example the discussion in~\cite{Kokkotas:1995av} and~\cite{Andersson:1995ez}. Thus, it is assumed that during the  merger of the NSs these modes are excited and  partially trapped inside the photon-sphere.  After some time, they leak outside with approximately the same frequency of the standing waves. The frequency of the GW echo is therefore determined by the eigenmodes of the photon-sphere trap, and is not related to the frequency of the GW emission during the inspiral. }

Most of the contribution to the integral in Eq.~\eqref{eq:tau_echo} comes from the strange star interior and
for the two considered models we obtain that the lowest  frequencies  are of the order of tens of kHz. In particular, for the last stable massive stars, corresponding to the tips of the SS1 and SS2 curves in Fig.~\ref{fig:MvsR}, we obtain $\omega_\text{1,echo}\simeq 17$ kHz and $\omega_\text{2,echo}\simeq 27$ kHz, respectively. Values of the bag constant \red{lying between $B_1$ and $B_2$} lead to intermediate values of the echo frequency. 

\section{ Conclusion}\label{sec:Conclusions}
We have  examined the possibility that a strange star has been  produced in the GW170817 merging event  and has  emitted a GW echo. 
Considering extreme strange star models   having a speed of sound equal to $1$, we have obtained that the most compact structures do cross the photon-sphere line, which is a necessary condition for producing GW echoes. However, the considered models do not approach the  Buchdahl's limit line  corresponding to  $R_B = 9/4 M$, which would lead to a GW echo emission at a frequency close to the values estimated in~\cite{Pani:2018flj} and thus approaching the frequency reported in~\cite{Abedi:2018npz}.
With our model the typical frequencies are of the order of $10$ kHz. 

 The basic reason of the discrepancy between our results and those of~\cite{Pani:2018flj}  is that strange quark matter is self-bound, but is not incompressible. Incompressible matter is characterized by a superluminal (actually infinite) speed of sound. In our approach we have instead assumed a speed of sound equal to the speed of light. In this case it is still possible to cross the photon-sphere line,  but the star cannot be too compact because at that point gravitational effects are large, leading to the gravitational collapse.   This leads to the typical behavior depicted in Fig.~\ref{fig:MvsR}, with the last stable compact configurations  close to the photon-sphere line.

We have neglected the stellar rotation and possible temperature effects on the EoS.
Regarding the stellar rotation, we have solved the TOV's equations assuming a static stellar model. However, including rotation it is expected to slightly change the GW echo frequency, see for example the estimates reported in~\cite{Pani:2018flj}.  Those estimates apply to the present model for the basic reason that strange stars are hardly deformable. 
Regarding the temperature effects, one should compare the expected temperatures produced in the NSs merging with the corresponding quark chemical potentials. Since in strange stars the quark chemical potential is of the order of hundreds of MeV, it seems unlikely that such a high temperature scale is produced in the merging or in the post-merger environment. 

We have restricted our analysis to strange stars, but different exotic ultracompact star models have been proposed, including boson stars~\cite{Wheeler:1955zz, Colpi:1986ye, Jetzer:1991jr}, see~\cite{Carignano:2016lxe, Brandt:2018bwq} for recent studies, and the  so-called Q-stars~\cite{Bahcall:1989ff}, both having a similar self-bound EoS. Whether they are sufficiently compact to approach the  Buchdahl's limit line is a topic that  will be considered  in a future work.

An interesting  possibility  is that the strange star produced by the  merging of NSs is in the unstable branch.  Since  stars in the unstable branch are more  compact than stable stars, they may lead to GW echoes at lower frequencies. In this case the star  would quickly collapse to a black hole, but it might have enough time to produce a GW echo signal. The estimated time for  NS collapse to black hole is of the order of the ms~\cite{1991A&A...252..651G, Shibata:2003iw, Baiotti:2004wn}, and it strongly depends on how far from equilibrium is the initial stellar configuration.
\red{A delayed collapse, on  timescales of  $10-100$ ms, is obtained for differentially rotating stars, see for example~\cite{Shapiro:2000zh}, and for stiff EoSs~\cite{Hotokezaka:2013iia}. We are not aware of any simulation of merging NSs leading to the formation of an unstable strange star, however, since the EoS in~\eqref{eq:EoS_quark_matter}  is extremely stiff, it may lead   to collapsing times of the order of $100$ ms or more. In this case, the collapsing time could be longer than $\tau_\text{echo}$, thus allowing, at least in principle, the emission of GW echoes at lower frequencies than those obtained in the present work. Note that for  realistic estimates of the echo timescale one should evaluate Eq.~\eqref{eq:tau_echo}  considering that the density and the pressure of the collapsing ultracompact star change with time.}
\acknowledgments
We thank Valeria Ferrari and Viviana Fafone for useful discussions.  

%\bibliography{BIB}

\end{document}